\documentclass[twocolumn,pra,showpacs,amsmath,amssymb]{revtex4}
\usepackage{graphicx}
\usepackage{amssymb,amsthm,amsfonts,amstext}
\usepackage{url}
\def\be{\begin{equation}}
\def\ee{\end{equation}}
\def\bea{\begin{eqnarray}}
\def\eea{\end{eqnarray}}
\def\bma{\begin{mathletters}}
\def\ema{\end{mathletters}}

\def\0{\overline{0}}

\def\q0{\underline{0}}

\def\H{{\cal H}}

\def\C{{\mathbb C}}
\def\id{{\mathbb I}}

\def\H{{\cal H}}

\def\R{\mathbb{R}}
\def\N{\mathbb{N}}
\def\Z{\mathbb{Z}}

\def\tr{\mbox{tr}}
\def\one{\leavevmode\hbox{\small1\normalsize\kern-.33em1}}

\def\bra#1{\langle#1|} \def\ket#1{|#1\rangle}
\def\braket#1#2{\langle#1|#2\rangle}

\def\proj#1{\ket{#1}\!\bra{#1}}

\newtheorem{theo}{Theorem}
\newtheorem{defin}[theo]{Definition}

\newtheorem{lemma}[theo]{Lemma}

\def\id{{\mathbb I}}
\def\tr{\mbox{tr}}

\begin{document}

\title{Sequential Strong Measurements and Heat Vision}

\author{Miguel Navascu\'es and
        David P\'erez-Garc\'ia}
\affiliation{Facultad de Matem\'aticas, Universidad Complutense de Madrid}

\begin{abstract}
We study scenarios where a finite set of non-demolition von-Neumann measurements are available. We note that, in some situations, repeated application of such measurements allows estimating an infinite number of parameters of the initial quantum state, and illustrate the point with a physical example. We then move on to study how the system under observation is perturbed after several rounds of projective measurements. While in the finite dimensional case the effect of this perturbation always saturates, there are some instances of infinite dimensional systems where such a perturbation is accumulative, and the act of retrieving information about the system increases its energy indefinitely (i.e., we have `Heat Vision'). We analyze this effect and discuss a specific physical system with two dichotomic von-Neumann measurements where Heat Vision is expected to show.
\end{abstract}

\maketitle

Classically, the ability to extract information from a physical system is only constrained by our technological means: ideally, a Maxwell demon could have a record of all the relevant variables of a given experiment, that could be actualized regularly without disturbing the dynamics of the system. In contrast, in quantum mechanics, measurement is an active process that inevitably modifies the system under observation. It is precisely the change the system faces during such a process that prevents us from measuring two non-commuting observables simultaneously.

Among the family of Positive Operator Valued Measures (POVMs) used to model the measurement process in quantum mechanics, projective or von Neumann measurements stand out due to their special properties. Von Neumann measurements do not only retrieve information, they also establish a property on the system under observation: once a measurement is made and a result obtained, any subsequent measurement will always output the previous value. Moreover, repeated measurements of the system will not modify its state any further. Von Neumann measurement are thus the closest analog of non-perturbative classical measurements, in the sense that they do not introduce external noise: any perturbation a system may experience is just a result of the state's non-definiteness of the property we want to estimate.

This last feature, the fact that consecutive repetitions of von Neumann measurements do not alter the physical state of the system, is the basis for the Zeno effect \cite{zeno}, where the evolution of a quantum system is literally frozen by a quasi-continuous measurement process. The Quantum Zeno effect has been shown recently to serve as a cooling mechanism in certain experimental situations \cite{cool}. On the other hand, from a tomographic point of view the Zeno effect is not very interesting, since the outcomes of all measurements performed are due to be the same.

In this work, we are interested in the compromise between the potential of von Neumann measurements for tomography and the perturbation of the quantum system over which they are implemented. We have just seen that allowing only one type of measurement leads to a very poor dynamics and not-very-informative tomography. This motivates us to study physical scenarios where a limited (finite) number of projective measurements with a finite number of outcomes each are available.

Let us first explore how the fact that we have several measurements to choose from can suppose a huge difference when we want to do tomography.

\vspace{5pt}

\noindent \emph{Quantum tomography}

In a classical system, the ability to perform two different measurements of $d$ outcomes each can only retrieve, at most, $2\log_2(d)$ bits of information. Indeed, if we measure the height of an individual and then its weight, we shall not expect to gather more information by measuring its height one more time. However, in the quantum world, two measurements $+,-$ will not commute in general. Given three instants of time $t_0\leq t_1\leq t_2$, this can lead to `paradoxical' situations where the output of measurement $+$ at time $t_0$ may differ from the output at time $t_2$ if there has been an intermediate measurement of $-$ at time $t_1$.

Sequential measurement schemes have already proven useful in the weak measurement scenario \cite{weak}, and could play an important role in future experimental tests of Quantum Mechanics \cite{fritz}. However, the possible applications of consecutive `strong' non-commuting measurements for tomography seem to be absent from scientific literature. This is certainly peculiar if one takes into account that, giving two non-commuting projectors $P,P'$, the operator space spanned by the POVM elements $\{P,PP'P,P'PP',PP'PP'P,...\}$ describing consecutive measurements may well be infinite dimensional, and so the corresponding estimated probabilities can give us access to an infinite number of parameters characterizing the state of our system. The current experimental credo, though, does not echo this simple observation: even when non-demolition interactions are available, systems are typically discarded after a single-shot measurement (e.g., in Wigner function estimation \cite{wigner}). Exploring many non-commuting properties of an unknown quantum state thus involves an unnecessarily large number of system initializations.

The previous ideas can be better illustrated by referring to a specific physical system.

Consider a spin $1/2$ neutral particle in a rectangular box of sizes $L\times L_0\times L_0$, with one of its vertices situated at the origin of coordinates. We will assume that $L_0\ll L$, so this system can be regarded as a one-dimensional object subject to the potential

\be
\begin{array}{c|l}V(x)=& 0,\mbox{ for } 0<x< L,\\ &\infty, \mbox{ otherwise.}\end{array}
\ee

The hamiltonian describing the evolution of such a particle is thus

\be
H=\frac{p^2}{2m}+ V(x),
\label{energy}
\ee

\noindent where $m$ denotes the mass of the particle. The eigenvectors of this system are $\{\ket{n}:n\in \N^+\}$ with $\braket{x}{n}=:\Psi_n(x)=\sqrt{\frac{2}{L}}\sin(\frac{\pi nx}{L})$, each with associated energy $E_n=\pi^2n^2/2mL^2$. The spin degree of freedom of the particle can be modeled through a 2-dimensional Hilbert space $\C^2$. Along these pages, the three pairs of kets $\{\ket{0},\ket{1}\}$, $\{\ket{+},\ket{-}\}$ and $\{\ket{+i},\ket{-i}\}$ will represent the eigenstates of the Pauli matrices $\sigma_x$, $\sigma_y$ and $\sigma_z$, respectively. It follows that the Hilbert space $\H\otimes \C^2$ where this particle lives can be then expanded in the basis $\{\ket{n,\pm i}:n\in\N^+\}$. As usual, quantum states in this scenario will be regarded as positive semidefinite normalized elements of $S_1=\{A:\tr(|A|)<\infty\}$, the set of trace-class operators. Not to be confused with the class of operators $S_2=\{A:\tr(AA^\dagger)<\infty\}$, that will also play an important role shortly.

Suppose now that our technology allows performing almost instantaneous von Neumann spin measurements along the $\hat{u}_z$ direction over such a particle without affecting its canonical degrees of freedom. Defining $\ket{\phi^{\pm}(x)}\equiv \cos(kx)\ket{0}\pm \sin(kx)\ket{1}$, then, the two von Neumann measurements $+$ and $-$ associated to the projectors

\be
F^{\pm}=\int \proj{x}\otimes \proj{\phi^{\pm}(x)}dx
\label{projectors}
\ee

\noindent can be physically realized by applying a magnetic field $\pm\vec{B}$ along the $\hat{u}_y$ direction with an intensity varying linearly with the $x$ coordinate \footnote{This can be implemented by creating a constant gradient field of the form $\vec{B}=-by\hat{u}_x-bx\hat{u}_y$ by means of a Maxwell coil \cite{maxwell}. Since the particle's $Y$ coordinate is null, the effective magnetic field will be $\vec{B}=-bx\hat{u}_y$.}, measuring the spin and then applying the inverse field $\mp \vec{B}$. Indeed, if $\left. \vec{B}\right|_{y,z=0}=-(bx)\hat{u}_y$ and the magnetic interaction is mediated through a hamiltonian $H_{s}=-\mu \vec{B}\cdot \vec{\sigma}$, one can check that

\be
F^{\pm}=e^{\mp iH_{s}\Delta t}\left(\id\otimes\proj{0}\right)e^{\pm iH_{s}\Delta t},
\ee

\noindent provided that we switch the field $\vec{B}$ for a time $\Delta t=\frac{k}{\mu b}$. This time will have to be very short (and so the magnetic density $b$ will have to be very strong) if we want to neglect the evolution of the system due to the main term (\ref{energy}) during $\Delta t$.

Module technicalities, we are thus able to implement two different von Neumann measurements over our system. We will call the outputs of such measurements $0$ and $1$ when the related projectors are $F^\pm$ and $\id-F^\pm$, respectively.

We will next show that the statistical analysis of repeated measurements allows reconstructing the probability density $\rho(x)$ of the particle inside the box. Before proceeding, note that we are not making any assumption on the initial coupling between the spin of the particle and its canonical degree of freedom (i.e., they could be classically correlated, or even entangled).

Suppose that we measure $+$ and $-$ alternatively $N$ times. For any $\vec{a}\in \{0,1\}^N$, define the probabily $P(\vec{a}|+-+-...)$ of obtaining outcomes $(a_1,a_2,a_3,...)$ during a sequential performance of the measurements $+,-,+,$... Then it is easy to check that

\begin{eqnarray}
&P(0,0,0,...|+-+...)+P(1,1,1,...|+-+...)=\nonumber\\
&=\left\langle \cos^{2(N-1)}(2kx)\right\rangle.
\label{tomography}
\end{eqnarray}

\noindent Using the fact that $\cos^{2M}(\theta)$ is a linear combination of the functions $\{\cos(2j\theta):j=0,...,M\}$, we can therefore estimate the values $\{\langle\cos(4kjx)\rangle:j=0,...,N-1\}$ from the statistical analysis of $N$ repeated measurements. If $k\leq \frac{\pi}{4L}$ we can then extend $\rho(x)$ to an even function defined in $[-\pi/4k,\pi/4k]$ and use our statistical knowledge to infer the coefficients $\{c_j\}_j$ of the Fourier expansion

\be
\rho(x)=\sum_{j=0}^\infty c_j \cos(4kjx), \mbox{ for } x\in [0,L].
\ee

\noindent At first glance, one could argue that, since the probabilities appearing in eq. (\ref{tomography}) decrease exponentially faster with $N$, it would require a large number of samples to estimate them. However, a proper calculation shows that, as long as $\rho(x)$ has a finite slope at $x=0$, the right-hand side of eq. (\ref{tomography}) decreases as $O(1/N)$.

We have just proven that, as we repeat measurements $+$ and $-$, we obtain more and more useful information about our initial state. However, as we will see soon, this information comes at a price.

\vspace{5pt}

\noindent \emph{Heat Vision}

As we said at the beginning, we are interested in how the state of the system is affected by the measurement process. Of course, one would expect such a state to depend on the implemented \emph{measurement strategy}, the process by which we choose which measurement to apply at a given time.

Any scenario where a limited number of von Neumann measurements are available can be studied by analyzing the statistics that result from randomly applying one measurement or the other. It is thus legitimate to consider the behavior of the system under \emph{independent random measurement strategies}, where the probability $p_x>0$ of performing measurement $x$ is the same on each round. We will see that the overall effect of random measurement strategies can have a very different nature depending on whether the dimension of the underlying Hilbert space is finite or infinite.

Suppose, indeed, that we perform with probability $p_x$ a measurement $x\in\{1,2,...,s\}$ defined by the complete set of projectors $\{F^x_a:a=1,2,...,d\}\subset B(\H)$, for some Hilbert space $\H$. The action of the resulting map $\Omega$ over an initial state $\sigma$ would then be given by

\be
\Omega(\sigma)=\sum_{x,a}p_xF^x_a\sigma F^x_a.
\label{paradigma}
\ee

\noindent Note that we can see $\sigma=\sum_{i,j}\ket{i}\bra{j}\in S_2$ as an element of $\H\otimes \H$ via the isomorphism \footnote{Indeed, notice that $\tr(\sigma^\dagger \sigma)=\braket{\sigma}{\sigma}$.} $\sigma\to\ket{\sigma}=\sum_{ij}\sigma_{ij}\ket{i}\ket{j}$. $\Omega$ can then be regarded as a superoperator $\overline{\Omega}\in B(\H\otimes \H)$, given by

\be
\overline{\Omega}=\sum_{x,a}p_xF^x_a\otimes (F^x_a)^*.
\ee

\noindent Seen as an operator, $\overline{\Omega}$ is both hermitian and positive semidefinite. Moreover, since this map is also unital, its operator norm will be upperbounded by 1 \cite{channel}. It follows that the spectrum of $\overline{\Omega}$ is in $[0,1]$ and so the limit $\lim_{N\to \infty}\overline{\Omega}^N$ exists and is equal to $\Pi_1$, the projector onto the space of eigenvectors of $\overline{\Omega}$ with eigenvalue 1. Note that such a limit does not depend on the initial probabilities $p_x$, as long as all of them are strictly positive. This is so because $\Pi_{1}$ is just the (possibly null) projector onto the intersection of the spaces $\H_x=\overline{\mbox{span}}\{\ket{\phi}:\sum_a F^x_a\otimes (F^x_a)^*\ket{\phi}=\ket{\phi}\}$.


Since in finite dimensions $S_2$ and the set of trace-class operators coincide, we can conclude that repeated applications of the mapping $\Omega$ will bring any quantum state $\sigma$ to a limiting state $\Omega^{\infty}(\sigma)\in S_1$. That is, even though the system may experience some perturbations at the beginning of the measurement process, given some time it will stabilize into a steady state. The finite dimensional case is thus resemblant of the quantum Zeno effect, i.e., repeated applications of an independent random measurement strategy will `freeze' the state with respect to some types of environmental noise.

In infinite dimensional systems, though, the norms $\|\cdot\|_1$ and $\|\cdot\|_2$ are not equivalent, so the convergence of the sequence of vectors $(\overline{\Omega}^N\ket{\sigma})$ in $\H\otimes \H$ does not necessarily imply the convergence of the sequence of states $(\Omega^N(\sigma))$ in $S_1$. In particular, it could happen that $\lim_{N\to\infty}\overline{\Omega}^N\ket{\sigma}=0\in S_2$.

As shown in Appendix \ref{nonconvergence}, if $(\Omega^N(\sigma))$ does not converge in $S_1$, such a sequence of quantum states will not admit a finite dimensional approximation. Moreover, the sequence of mean values $\tr(\Omega^N(\sigma)H)$ can be shown to diverge for any 0-band hamiltonian $H$ iff $\Omega^N(\sigma)$ does not converge in $S_1$. In such a situation, the action of gathering information about the system will then increase its energy up to arbitrarily high values. We will hence call such an effect `Heat Vision', in analogy with Superman's famous ability to induce heat with the power of his stare \cite{superman}.

We will next prove that, remarkably, our previous example exhibits Heat Vision for any initial state $\sigma\in S_1$ we place as an input.

First, since Heat Vision does not depend on the actual weights we assign to each measurement, we can assume w.l.o.g. that we perform any of the two measurements with probability 1/2. The measurement channel $\Omega$ is thus equal to

\be
\Omega(\sigma)=\frac{1}{2}\{\sum_{j=+,-}F^j\sigma F^j+(\id-F^j)\sigma(\id-F^j)\}.
\label{type1}
\ee

As shown in Appendix \ref{derivation}, for any state $\sigma$,

\be
\lim_{N\to\infty}\tr\{[\Omega^N(\sigma)]^2\}=0.
\label{zero_conv}
\ee

\noindent Clearly, the values $(\|\Omega^N(\sigma)-0\|_1)$ do not tend to 0, and so the sequence of states $(\Omega^N(\sigma))$ does not converge in trace norm. Also, equation (\ref{zero_conv}), together with the R\'enyi inequality $S(\chi)\geq -\log_2(\tr\{\chi^2\})$, valid for any normalized quantum state, implies that the von Neumann entropy of $\Omega^N(\sigma)$ tends to infinity.

As we mentioned earlier, the fact that $(\Omega^N(\sigma))$ does not converge in $S_1$ implies that the energy of the system will diverge for any input state, as long as the hamiltonian describing the system has no energy bands in its spectrum. Such is the case of the hamiltonian described by eq. (\ref{energy}), and, indeed, one can check that the kinetic energy of a quantum state after $N$ applications of the channel $\Omega$ obeys the formula

\be
E^{(N)}=E^{(0)}+\frac{k^2}{m}N.
\ee

\noindent That is, on average, the temperature of the system will grow linearly with the number of measurements performed.

Let us briefly recapitulate what is happening here: in order to perform measurement $\pm$, first we have to apply a strong magnetic field $\pm\vec{B}$ for a time. It is not surprising then that the energy of the system increases when we perform such a change. However, after the spin measurement we will apply the opposite field, $\mp\vec{B}$, thus inverting the previous unitary process. The only reason why the energy of the system (or for the same sake, the state of the system) changes is thus that we are measuring a single qubit in between! Indeed, the system is engineered in such a way that the entropy associated to such a spin measurement accumulates and accumulates in the canonical degree of freedom of our particle until the setup cannot stand more energy. This behavior is to be compared to the finite dimensional case, where it is always possible to find states of entropy 1 that are invariant under the action of any two dichotomic measurements \footnote{In a finite dimensional system, any pair of projectors $F^+$ and $F^-$ can be simultaneously $2\times 2$-block-diagonalized, i.e., there exists a basis where $F^+=\oplus_n F_n^+,F^-=\oplus_n F_n^-$, with $F_n^+,F_n^-\in M_{2\times 2}$ (\textbf{Jordan}, 1875). Any state of the form $\sigma=0_{2\times 2}\oplus ...\oplus \frac{\id_2}{2}\oplus ...\oplus 0_{2\times 2}$ thus satisfies $F^\pm\sigma F^\pm+(\id-F^\pm)\sigma (\id-F^\pm)=\sigma$.}.

But Heat Vision can manifest in even more extreme ways: in Appendix \ref{extreme} we describe systems with 5 dichotomic observables where the purity of any initial state $\sigma$ subject to $N$ sequential measurements is bounded by $\lambda^{2N}$, where $\lambda<1$ is independent of $\sigma$. The extropy of such states will thus increase at least linearly in $N$. The measurements involved, though, are quite abstract, and most likely impossible to implement in any present laboratory.

\begin{figure}
  \centering
  \includegraphics[width=8 cm]{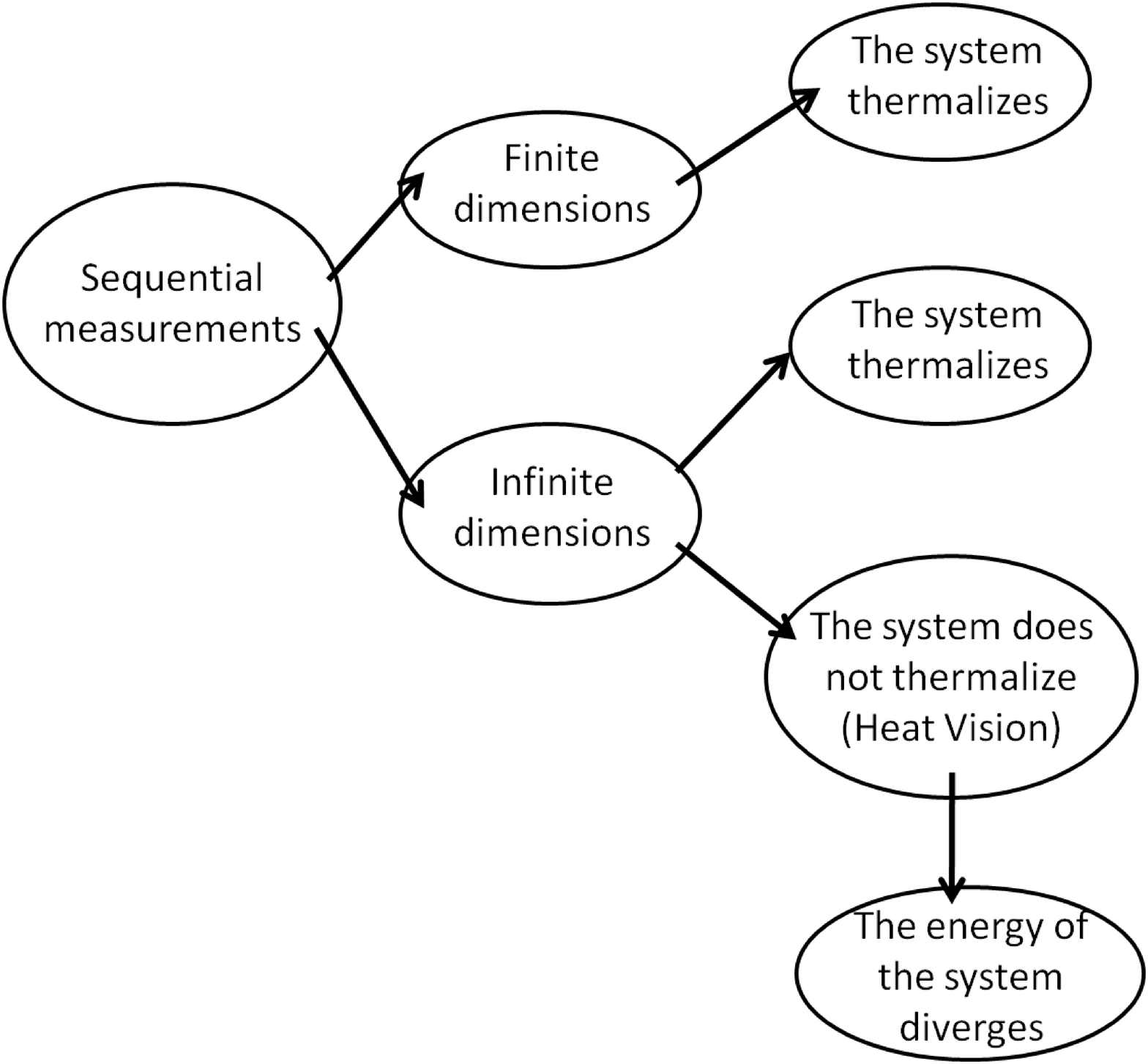}
  \caption{Dynamics of a quantum system subject to repeated von Neumann measurements.}
  \label{dynamics}
\end{figure}

By showing that Heat Vision does emerge (at least, theoretically) in some physical systems, now we have a clear picture of how quantum systems evolve through sequential von Neumann measurements. The conclusions are illustrated in Figure \ref{dynamics}, where the finite and infinite dimensional case are differentiated.

It is tempting to think that Heat Vision arises in some infinite dimensional systems and not in finite dimensional ones just because in the former we can extract an infinite amount of information about the initial state of the system. However, in Appendix \ref{non_heat} we describe a system with two dichotomic von Neumann measurements where the statistical analysis of sequential observations retrieves infinitely many parameters, but Heat Vision never shows up. Heat Vision is thus not equivalent to the possibility of accessing an infinite amount of information, but an independent property of the system under study.

Finally, we would like to point out that both tomography with strong sequential measurements and Heat Vision can be experienced with current technology. Indeed, consider a regular ion trap setup with just one ion. Then two internal states of such an ion could play the role of the spin in the system described above, while the displacement of the ion along the trap could account for the canonical degree of freedom. `Spin' measurements in this scenario can be performed in the standard way, i.e., exciting one of the internal levels with a laser and counting the number of emitted photons. Analogously, an interaction of the form $H_s$ can be induced by a laser beam in standing wave configuration \cite{review}. In a usual ion trap setup, the former dispositions would effectively implement the measurements (\ref{projectors}) over a particle subject to a harmonic potential; in order to recreate the square potential, a trap of the form \cite{square} can be used. Due to the Brans-Dicke approximation, though, the whole experiment must be conducted in the low temperature regime (up to $\sim 1$K).

\vspace{5pt}

\noindent \emph{Conclusion}

In this paper we have studied the use and effect of repeated von Neumann measurements. We have pointed out an extreme scenario where the statistical analysis of binary outcomes of sequential measurements allows estimating the full probability density of a trapped particle. We have also shown how sometimes the action of alternating measurements can lead to a non-convergent dynamics in infinite quantum dimensional systems. This phenomenon always comes together with an unbounded energy increase (hence its name `Heat Vision'), and can be observed experimentally in current ion trap setups. Is this the end of the story? Dreaming on, one could conceive a new architecure for quantum computing based on sequential strong measurements. In this model, the user could perform a (small) number of non-commuting dichotomic measurements over a continuous variable system, and computations would be carried out by deciding which measurement to perform at every step. Although still vague, we hope to explore this idea in future communications.


%

The authors would like to thank A. Retzker for suggesting experimental implementations. This work was supported by the Spanish grants I-MATH, MTM2008-01366, S2009/ESP-1594 and the European projects QUEVADIS.

\begin{appendix}
\clearpage

\section{Convergence in $S_1$}
\label{nonconvergence}

\begin{theo}
\label{equiva}
Let $\sigma^{(N)}$ be a sequence of normalized quantum states. Then, the following conditions are equivalent:

\begin{enumerate}
\item $\lim_{N\to\infty}\sigma^{(N)}$ exists in $S_1$.

\item For some (and thus for all) arbitrary countable orthonormal basis $\{\ket{n}\}$ of $\H$, the limits

\be
c_{n,m}=\lim_{N\to\infty}\tr\{\sigma^{(N)}\ket{n}\bra{m}\}
\label{coefic}
\ee

\noindent exist and are such that

\be
\sum_{n=0}^\infty c_{n,n}=1.
\label{complete}
\ee

\end{enumerate}
\end{theo}

\begin{proof}

Suppose that 1) is true. Then, there exits an element $\hat{\sigma}\in S_1$ such that $\lim_{N\to\infty}\|\hat{\sigma}-\sigma^{(N)}\|_1=0$. We remind the reader that, for any self-adjoint operator $A$,

\be
\|A\|_1=\sup_{\id\geq X\geq -\id} \tr\{A\cdot X\}.
\ee

\noindent Now, let $\{\ket{n}\}$ be any orthonormal basis for $H$. The operator $\proj{n}$ satisfies $\id\geq \pm\proj{n}\geq -\id$, so

\be
\|\hat{\sigma}-\sigma^{(N)}\|_1\geq |\tr\{\proj{n}\hat{\sigma}\}-\tr\{\proj{n}\sigma^{(N)}\}|.
\ee

\noindent It follows that the limit $\lim_{N\to\infty}\tr\{\sigma^{(N)}\proj{n}\}$ exists and is equal to $\tr\{\hat{\sigma}\proj{n}\}$. Analogously, from the relations

\begin{eqnarray}
&&\id\geq \ket{n}\bra{m}+\ket{m}\bra{n}\geq-\id,\nonumber\\
&&\id\geq i(\ket{n}\bra{m}-\ket{m}\bra{n})\geq-\id,
\end{eqnarray}

\noindent it can be shown that $\lim_{N\to\infty}\tr\{\sigma^{(N)}\ket{n}\bra{m}\}$ exists as well. Finally, $\id\geq\pm\id\geq -\id$, which means that $\|\hat{\sigma}-\sigma^{(N)}\|_1\geq |\tr\{\hat{\sigma}-\sigma^{(N)}\}|$. Since $\tr(\sigma^{(N)})=1, \forall N$, we have that $\tr(\hat{\sigma})=1$, and so $\sum_n c_{n,n}=\tr\{\sum_n\proj{n}\hat{\sigma}\}=\tr(\hat{\sigma})=1$, and 2) is proven true.

Conversely, suppose that 2) is true, and consider the operator

\be
\hat{\sigma}\equiv\sum_{n,m=0}^\infty c_{n,m}\ket{m}\bra{n}.
\ee

\noindent

This operator is bounded. Indeed, let $\ket{v},\ket{w}\in \mbox{span}\{\ket{n}\}$. Then,

\be
|\bra{v}\hat{\sigma} \ket{w}|=\lim_{N\to\infty}|\tr\{\sigma^{(N)}\ket{w}\bra{v}\}|\leq \sqrt{\braket{w}{w}\braket{v}{v}}.
\ee

\noindent Likewise, it can be shown that $\bra{v}\hat{\sigma} \ket{v}\geq 0$ for all $\ket{v}\in \mbox{span}\{\ket{n}\}$, i.e., $\hat{\sigma}\geq 0$. Moreover, by equation (\ref{complete}) $\tr\{\hat{\sigma}\}=1$, so $\hat{\sigma}\in S_1$.

Let $P_K\equiv\sum_{n=0}^K\proj{n}$. Then equation (\ref{complete}) implies that, for any $\epsilon>0$, there exist $K,M$ such that $\tr(\sigma^{(N)}P_K)\geq 1-\epsilon,\forall N\geq M$. Applying twice the relation \cite{renner}

\be
\|\rho-P_K\rho P_K\|_1\leq 2\sqrt{\tr(\rho)\{\tr(\rho)-\tr(\rho P_K)\}},
\ee

\noindent valid for any $\rho\geq 0\in S_1$, we have that

\be
\|\sigma^{(N)}-\hat{\sigma}\|_1\leq 4\sqrt{\epsilon}+\|P_K\sigma^{(N)}P_K-P_K\hat{\sigma}P_K\|_1,
\label{convS1}
\ee

\noindent for $N>M$. Note that the last term of Eq. (\ref{convS1}) tends to 0 as $N$ tends to infinity (because we are evaluating the trace distance between two $K+1\times K+1$ matrices that converge entry-wise). It follows that $\lim_{N\to\infty}\|\sigma^{(N)}-\hat{\sigma}\|_1\leq 4\sqrt{\epsilon}$. Since $\epsilon$ was arbitrary, we conclude that $\lim_{N\to\infty}\|\sigma^{(N)}-\hat{\sigma}\|_1=0$, and so $(\sigma^{(N)})$ converges in $S_1$.

\end{proof}

In the particular case where $\sigma^{(N)}=\Omega^N(\sigma)$, for some initial state $\sigma$ and some channel $\Omega$ with $\id\geq\overline{\Omega}\geq 0$, the existence of the limits (\ref{coefic}) is automatic, since $\overline{\Omega}^N$ converges in $B(\H\otimes \H)$ and $\ket{n}\ket{m}^*\in \H\otimes \H$. This implies that convergence in $S_1$ in that case is equivalent to the existence of a basis $\{\ket{n}:n\in \N\}$ such that $\lim_{K\to\infty}\lim_{N\to\infty} \tr(P_K\Omega^N(\sigma))=1$ (note the order of the limits). If the latter is the case, then $\sigma^{(N)}$ can always be described by a finite dimensional system, i.e., for sufficiently high $K$, we can approximate $\sigma^{(N)}$ by the state $P_K\sigma^{(N)}P_K\in B(\C^{K+1})$, for all $N$.

In order to establish a connection between energy and convergence in $S_1$ in realistic scenarios, we will have to restrict the usual definition of hamiltonian.

\begin{defin}\textbf{0-band energy operator}\\
Let $E$ be a self-adjoint operator acting over an infinite dimensional (separable) Hilbert space $\H$. We will say that $E$ is a 0-band energy operator iff

\begin{enumerate}
\item The spectrum of $E$ is discrete.

\item For any $\bar{E}\in \R$, there is only a finite number of linearly independent eigenvectors of $E$ with eigenvalues less or equal than $\bar{E}$.

\end{enumerate}

\end{defin}

Examples of 0-band energy operators are the harmonic oscillator, the double-well potential, a particle in a box... and, more generally, the hamiltonian of any finite number of particles subject to a potential that can be bounded from below by a harmonic trap. Since most experiments in Physics are not performed in the open air, but inside closed chambers, 0-band energy operators provide a very good dynamical description of those quantum systems accessible in the lab.

The next lemma relates convergence in $S_1$ with energy considerations.

\begin{lemma}
\label{conv_ener}
Let $(\sigma^{(N)})$ be a sequence of normalized quantum states such that $\lim_{N\to\infty} \sigma^{(N)}$ exists in $S_2$. Then, $\lim_{N\to\infty}\sigma^{(N)}$ exists in $S_1$ iff, for some 0-band energy operator $E$ $\lim_{N\to\infty}\tr\{E\sigma^{(N)}\}$ exists.

\end{lemma}

\begin{proof}
Suppose that the sequence $(\sigma^{(N)})$ does not converge in $S_1$, and let $E$ be an arbitrary 0-band energy operator $E=\sum_{n=0}^\infty E_n\proj{n}$, where $\{\ket{n}:n\in\N\}$ are a basis of eigenvectors of $E$, with energies $E_0\leq E_1\leq E_2\leq...$. Let $P_K$ be the projector $P_K\equiv\sum_{n=0}^K\proj{n}$. We will prove that there exists $1>\lambda>0$ such that, for any $\bar{E}\geq E_0$, $\lim_{N\to\infty}\tr\{E\sigma^{(N)}\}\geq (1-\lambda)E_0+\lambda\bar{E}$.

Indeed, let $K$ be such that $E_{K+1}\geq \bar{E}$. If $\lim_{N\to\infty}\sigma^{(N)}$ does not converge in $S_1$, by Theorem \ref{equiva} we have that $\lim_{N\to\infty}\tr\{\sigma^{(N)}P_K\}\leq\lim_{n\to\infty}\lim_{N\to\infty}\tr\{\sigma^{(N)}P_n\}=1-\lambda$, with $\lambda>0$ independent of $K$. On the other hand,

\be
E=\sum E_n\proj{n}\geq E_0P_K+ \bar{E}(\id-P_K),
\ee

\noindent so we arrive at

\be
\lim_{N\to\infty}\tr\{E\sigma^{(N)}\}\geq (1-\lambda)E_0+ \lambda\bar{E}.
\ee

Conversely, suppose that $(\sigma^{(N)})$ converges to the normalized state $\hat{\sigma}$. We will prove that there exists a 0-band energy operator $E$ such that $\tr\{\sigma^{(N)}E\}$ is bounded and $\lim_{N\to\infty}\tr\{\sigma^{(N)}E\}$ exists. Let $\{\ket{n}:n\in\N\}$ be any basis for $\H$ and define the probability distributions $p^{(N)}(n)\equiv \bra{n}\sigma^{(N)}\ket{n}$, $\hat{p}(n)\equiv \bra{n}\hat{\sigma}\ket{n}$. Also, let $K:\N\to \N$ be the mapping defined as

\be
K(s)\equiv \{\min K\geq 0: \sum_{n=0}^Kp^{(N)}(n)\geq 1-\frac{1}{2^s},\forall N\}.
\ee

\noindent It is important to note that $K(s)<\infty$ for all $s$. Indeed, suppose that $K(s)=\infty$, for some $s$. That would imply that, for any number $K$, there exists an $\tilde{N}$ such that $\sum_{n=0}^K p^{(\tilde{N})}(n)< 1-\frac{1}{2^s}$. Now, for $0<\epsilon<1/2^{s+1}$, choose $M$ such that $\|\hat{\sigma}-\sigma^{(N)}\|_1<\epsilon$, for all $N>M$, and choose $K$ such that

\begin{eqnarray}
&&\sum_{n=0}^Kp^{(N)}(n)>1-\frac{1}{2^s}+\epsilon, \forall N\leq M,\nonumber\\
&&\sum_{n=0}^K\hat{p}(n)>1-\frac{1}{2^s}+2\epsilon.
\end{eqnarray}

\noindent Then, for $N>M$, $|\sum_{n=0}^Kp^{(N)}(n)-\hat{p}(n)|\leq\|\sigma^{(N)}-\hat{\sigma}\|_1<\epsilon$. It follows that the expression $\sum_{n=0}^Kp^{(N)}(n)>1-\frac{1}{2^s}+\epsilon$ holds for all $N$, thus contradicting our initial claim.

We will differentiate two cases depending on the existence of the limit $\lim_{s\to\infty}K(s)$.

If $\lim_{s\to\infty}K(s)=\hat{K}<\infty$, then $\sum_{n=0}^{\hat{K}}p^{(N)}(n)=1$ for all $N$. We can thus simply define the 0-band energy operator $E=\sum_{n=\hat{K}+1}^\infty n\proj{n}$ and we would have that $\tr\{\sigma^{(N)}E\}=0<\infty$.

Suppose, on the contrary, that $\lim_{s\to\infty}K(s)=\infty$, and define the sets of natural numbers $I_0=[0,K(1)]$ and

\begin{eqnarray}
I_s\equiv && [K(s)+1,K(s+1)], \mbox{ if } K(s)+1\leq K(s+1)\nonumber\\
&& \emptyset, \mbox{ otherwise},
\end{eqnarray}

\noindent for $s\geq 1$. These sets are finite and disjoint, and satisfy $\cup_{s=0}^\infty I_s=\N$. Denoting the projector $\sum_{n\in I_s}\proj{n}$ as $P_s$, we thus have that the positive operator

\be
E=\sum_{s=0}^\infty \sqrt{2}^sP_s
\ee

\noindent is 0-banded.

\noindent Finally, notice that

\be
\sum_{n\in I_s}p^{(N)}(n)\leq\sum_{n=K(s)+1}^{\infty}p^{(N)}(n)\leq \frac{1}{2^s},
\ee

\noindent for $s\geq 1$. This implies that, for any $N$,

\be
E^{(N)}\equiv\tr(\sigma^{(N)}E)\leq 1+\sum_{s=1}^\infty\left(\frac{\sqrt{2}}{2}\right)^s=1+\frac{1}{\sqrt{2}-1},
\ee

\noindent i.e., the sequence $(E^{(N)})$ is bounded.

Now, define $Q_t=\sum_{s=0}^t P_s$. From all the above, it is clear that

\be
0\leq E^{(N)}-\tr(Q_t\sigma^{(N)} Q_tE)\leq \sum_{s=t+1}^\infty\left(\frac{\sqrt{2}}{2}\right)^s.
\ee

\noindent Likewise,

\be
0\leq \hat{E}- \tr(Q_t\hat{\sigma} Q_tE)\leq\sum_{s=t+1}^\infty\left(\frac{\sqrt{2}}{2}\right)^s,
\ee

\noindent where $\hat{E}\equiv \tr(\hat{\sigma}E)$. It follows that

\be
|\hat{E}-E^{(N)}|\leq |\tr(Q_t[\hat{\sigma}-\sigma^{(N)}]Q_t E)|+\frac{\sqrt{2}-1}{\sqrt{2}^t}.
\label{finprueba}
\ee

Taking the limit $N\to\infty$ and then $t\to\infty$, the right-hand-side of equation (\ref{finprueba}) vanishes, and thus $\lim_{N\to\infty}\tr(\sigma^{(N)}E)=\tr(\hat{\sigma} E)$.

\end{proof}

Let us make a final remark.

\begin{lemma}
\label{zeroes}
Suppose that the channel $\id\geq\overline{\Omega}\geq 0$ has the property that, for any quantum state $\sigma$, the sequence $(\Omega^N(\sigma))$ does not converge in $S_1$. Then, $\lim_{N\to \infty}\Omega^N(\sigma)=0$ in $S_2$, i.e., $\lim_{N\to\infty}\tr\{\sigma^{(N)}\ket{\phi}\bra{\psi}\}=0$ for any pair of states $\ket{\phi},\ket{\psi}$.
\end{lemma}

\begin{proof}
Suppose that there exists a basis $\{\ket{n}:n\in \N\}$ for the Hilbert space such that the coefficients $c_{n,m}=\lim_{N\to\infty}\tr(\Omega^N(\sigma)\ket{n}\bra{m})$ do not satisfy condition (\ref{complete}). Then, following the proof of Theorem \ref{equiva}, one could build an operator $\hat{\sigma}\geq 0\in S_1$ such that $\tr(\hat{\sigma})\leq 1$. Now, if $\tr(\hat{\sigma})\not=0$, then $\hat{\sigma}'\equiv\hat{\sigma}/\tr(\hat{\sigma})$ would be a quantum state such that $\Omega(\hat{\sigma}')=\hat{\sigma}'$, contradicting the main assumption. We thus have that $\tr(\hat{\sigma})=0$ which, together with $\hat{\sigma}\geq 0$ implies that $\hat{\sigma}=0$ (and so, $\bra{\phi}\hat{\sigma}\ket{\psi}=\lim_{N\to\infty}\tr\{\sigma^{(N)}\ket{\psi}\bra{\phi}\}=0$, for all $\ket{\psi},\ket{\phi}$).
\end{proof}

\section{Derivation of Equation (\ref{zero_conv})}
\label{derivation}

Viewed as a superoperator, the channel (\ref{type1}) can be seen equal to

\begin{eqnarray}
\overline{\Omega}=&&\frac{1}{2}(\id\otimes U)\int dxdy \proj{x}\otimes\proj{y}\otimes\nonumber\\
&&\otimes \left(M(x-y)\oplus M(x+y)\right)(\id\otimes U^\dagger).
\end{eqnarray}

\noindent where

\begin{eqnarray}
M(x)&=&\left(\begin{array}{cc}1&\cos[2kx]\\ \cos[2kx]&1\end{array}\right)\nonumber\\
&=&\cos^{2}[kx]\proj{+}+\sin^{2}[kx]\proj{-},
\end{eqnarray}

\noindent and

\begin{eqnarray}
U=&&\ket{+i,-i}\bra{0,0}+\ket{-i,+i}\bra{0,1}+\nonumber\\
&&+\ket{+i,+i}\bra{1,0}+\ket{-i,-i}\bra{1,1}.
\end{eqnarray}

\noindent It thus follows that

\begin{eqnarray}
\overline{\Omega}^N=&&\frac{1}{2}(\id\otimes U)\int dxdy \proj{x}\otimes\proj{y}\otimes\nonumber\\
&&\otimes\left(M(x-y)^N\oplus M(x+y)^N\right)(\id\otimes U^\dagger),
\end{eqnarray}

\noindent and one can then check that

\begin{eqnarray}
&&\tr\{\left[\Omega^N(\proj{n,\pm i})\right]^2\}=\nonumber\\
&&=\bra{n,\pm i,n,\mp i}\overline{\Omega}^{2N}\ket{n,\pm i,n,\mp i}=\nonumber\\
&&=\frac{1}{2}\int_{0}^L\int_{0}^L dxdy \Psi_n(x)^2\Psi_n(y)^2\cdot\nonumber\\
&&\cdot\left(\cos^{4N}[k(x-y)]+\sin^{4N}[k(x-y)]\right).
\end{eqnarray}

\noindent Taking into account that $\Psi_n(x)^2\Psi_n(y)^2\leq 4/L^2$ and changing to variables $x'=(x-y)/L,y'=(x+y)/L$, we have that

\begin{eqnarray}
&\tr\{\left[\Omega^{N}(\proj{n,\pm i})\right]^2\}\leq \varphi(N)^2:=\nonumber\\
&=2\int_{-1}^1 [1-|x|]\cdot[\cos^{4N}(kLx)+\sin^{4N}(kLx)]dx.
\end{eqnarray}

\noindent For any initial state $\sigma$, applying the Schwartz inequality one arrives at

\begin{eqnarray}
&&\tr\{\Omega^N(\sigma)\proj{n,\pm i}\}=\tr\{\sigma\Omega^N(\proj{n,\pm i})\}\leq\nonumber\\ &&\leq\sqrt{\tr(\sigma^2)}\sqrt{\tr\{\left[\Omega^{N}(\proj{n,\pm i})\right]^2\}}\leq \varphi(N),
\end{eqnarray}

\noindent with $\lim_{N\to\infty}\varphi(N)=0$. Note that this expression is independent of $n$: the occupation of each of the states $\{\ket{n,\pm i}\}$ tends uniformly to zero. In other words, the energy density distribution of a quantum state subject to sequential measurements $+$ and $-$ neither converges nor displaces, but flattens.

Now, Theorem \ref{equiva} in Appendix \ref{nonconvergence} states that any sequence $(\Omega^N(\sigma))$ does not converge in $S_1$ iff there exists some orthonormal basis $\{\ket{\psi_n}:n\in \N\}$ such that $\sum_n \lim_{N\to\infty}\bra{\psi_n}\Omega^N(\sigma)\ket{\psi_n}<1$. Moreover, by lemma \ref{zeroes} in the same Appendix, if such is the case for any initial state $\sigma$, then $\lim_{N\to\infty}\Omega^N(\sigma)$ tends to 0 in the $\|\cdot\|_2$ norm. It thus follows that

\be
\lim_{N\to\infty}\tr\{[\Omega^N(\sigma)]^2\}=0,
\ee

\noindent for all initial states $\sigma$.

\section{An example of unbounded extraction of information with no heat vision}
\label{non_heat}
Consider a separable Hilbert space $\H$, and let $\{\ket{n}:n\in \N\}$ be an orthonormal basis for $\H$. Then, we can define projector operators acting over $\H\otimes \C^2$ as

\be
G^{\pm}=\sum_{n=1}^\infty \proj{n}\otimes \proj{\phi^{\pm}(n)}.
\ee

\noindent The analog of equation (\ref{tomography}) follows straightforwardly, and so we can combine our projective measurements to estimate the mean values $\{\langle\cos(4kjn)\rangle:j\in \N\}$. Choosing $k$ irrational, the statistical analysis of repeated measurements can thus provide us with an infinite amount of information about the ocupation number distribution. However, let $\Omega$ be the channel that results when we randomly apply one measurement or the other. Then, for any initial quantum state $\sigma$, it can be shown that

\be
\lim_{N\to\infty}\Omega^N(\sigma)=\sum_{n=1}^\infty\bra{n}\tr_{\C^2}(\sigma)\ket{n} \proj{n}\otimes \id_2/2,
\ee

\noindent that is, the system does not exhibit heat vision in any case. Moreover, if the energy operator is diagonal in the $\{\ket{n}:n\in\N\}$ basis, the energy of the system does not even vary during the measurement process.

\section{Extreme Cases of Heat Vision}
\label{extreme}
An extreme case of heat vision can be found in the following system. Consider the group $G$ that results out of the free product \cite{freeproduct} of $\Z_2$ with itself $s$ times, i.e., $G=\overbrace{\Z_2*\Z_2*...*\Z_2}^{s \mbox{ times }}$, and take $l_2(G)$ to be our Hilbert space.

In this context, the left regular representation of an element $g\in G$ is defined as the unitary operator $\lambda(g):l_2(G)\to l_2(G)$ such that $\lambda(g)\ket{g'}=\ket{gg'}$, for any $g'\in G$ \cite{pisier}. The left regular representation of the generators $\lambda(g_i)$ thus satisfies $\lambda(g_i)^2=1,\lambda(g_i)=\lambda(g_i)^\dagger$. This implies that, for each generator $g_i$, there exists an associated projector $(\lambda(g_i)+\id)/2$, and so each $g_i$ defines a quantum dichotomic measurement. The channel $\Omega$ that results when we perform one of the $s$ measurements with probability $1/s$ can be written as

\be
\overline{\Omega}=\frac{1}{2}\left(\frac{1}{s}\sum_{i=1}^s \lambda(g_i)\otimes\lambda(g_i)^*+\id\right).
\ee

Define the operator $\Pi_i$ as the projection onto the subspace of $l_2(G)$ spanned by all the elements of $G$ that start with the symbol $g_i$, and note that $\lambda(g_i)=x_i+y_i$, with $x_i\equiv\lambda(g_i)\Pi_i, y_i\equiv\Pi_i\lambda(g_i)$. We have that

\be
\|\sum_{i=1}^s\lambda(g_i)\otimes \lambda(g_i)^*\|= \|\sum_{i=1}^s\lambda(g_i)\|\leq \|\sum_{i=1}^s x_i\|+\|\sum_{i=1}^s y_i\|,
\label{intermedio}
\ee

\noindent where in the first equality we have made use of Fell's absortion principle \cite{pisier}. On the other hand, for any two sets of operators $\{A_i\}$ $\{B_i\}$,

\be
\|\sum_i A_iB_i\|\leq \|\sum_i A_i A_i^\dagger\|^{1/2}\|\sum_i B_i^\dagger B_i\|^{1/2}.
\ee

\noindent Taking $(A_i=\id,B_i=x_i)$ and $(A_i=y_i,B_i=\id)$, we have that the last term of equation (\ref{intermedio}) is upperbounded by

\be
\sqrt{s}\left(\|\sum_i \Pi_i\|^{1/2}+\|\sum_i \Pi_i\|^{1/2}\right),
\ee

\noindent that, in turn, is upperbounded by $2\sqrt{s}$, since $\sum_i \Pi_i\leq \id$.

It follows that

\be
\|\overline{\Omega}\|\leq 1/2+1/\sqrt{s}.
\ee

\noindent The norm of $\overline{\Omega}$ as an operator in $l_2(G)\otimes l_2(G)$ is thus smaller than 1 whenever the number of measurements is greater than 4. This phenomenon can only occur in infinite dimensional systems, since, for any finite dimensional unital map $\overline{\omega}$ on $\C^d\otimes \C^d$, $\overline{\omega}\ket{\id_d}=\ket{\id_d}$, for $\ket{\id_d}\equiv\sum_{i=1}^d\ket{i,i}$, and so $\|\overline{\omega}\|=1$.

Suppose, then, that $s\geq 5$ and so $\|\overline{\Omega}\|=\lambda<1$, and let $\sigma\in S_1(l_2(G))$ be any arbitrary normalized quantum state, with $\sigma^{(N)}\equiv\Omega^N(\sigma)$. Our previous discussion implies that

\be
\tr\{[\sigma^{(N)}]^2\}\leq \lambda^{2N}\tr\{[\sigma]^2\}\leq \lambda^{2N}.
\label{purity}
\ee

\noindent That is, the purity of any initial state decreases exponentially with the number of applications of the channel, and so the system exhibits Heat Vision for any input state.

\end{appendix}

\end{document}